\documentclass[submission,copyright,creativecommons]{eptcs}
\providecommand{\event}{LSFA 2026} 

\usepackage{iftex}

\ifpdf
  \usepackage{underscore}         
  \usepackage[T1]{fontenc}        
\else
  \usepackage{breakurl}           
\fi

\usepackage{xspace}
\newcommand*{\cslib}{CSLib\xspace}
\newcommand{\grind}{\lstinline|grind|\xspace}

\usepackage{xcolor}
\usepackage{listings}
\usepackage{amssymb}

\definecolor{primary}{RGB}{0, 51, 102}    
\definecolor{secondary}{RGB}{230, 240, 255} 
\definecolor{accent}{RGB}{100, 100, 100}  

\definecolor{keywordcolor}{rgb}{0.7, 0.1, 0.1}   
\definecolor{tacticcolor}{rgb}{0.0, 0.1, 0.6}    
\definecolor{commentcolor}{rgb}{0.4, 0.4, 0.4}   
\definecolor{symbolcolor}{rgb}{0.0, 0.1, 0.6}    
\definecolor{sortcolor}{rgb}{0.1, 0.5, 0.1}      
\definecolor{attributecolor}{rgb}{0.7, 0.1, 0.1} 

\title{Computer Science as Infrastructure: the Spine of the Lean Computer Science Library (\cslib)}
\author{
Christopher Henson
\institute{Drexel University\\
Philadelphia, USA}
\email{\quad ch3473@drexel.edu}
\and
Fabrizio Montesi
\institute{FORM, University of Southern Denmark\\
Odense, Denmark}
\email{\quad fmontesi@imada.sdu.dk}
}
\def\titlerunning{The Lean Computer Science Library}
\def\authorrunning{C. Henson \& F. Montesi}
\begin{document}
\maketitle

\begin{abstract}
Following in the footsteps of the success of Mathlib -- the centralised library of formalised mathematics in Lean -- \cslib is a rapidly-growing centralised library of formalised computer science and software.
In this paper, we present its founding technical principles, operation, abstractions, and semantic framework. We contribute reusable semantic interfaces (reduction and labelled transition systems), proof automation, CI/testing support for maintaining automation and compatibility with Mathlib, and the first substantial developments of languages and models.
\end{abstract}

\section{Introduction}

The Lean proof assistant \cite{de_moura_lean_2015} and the Mathlib library \cite{the_mathlib_community_lean_2020} have brought formal methods closer to the mainstream mathematics community by offering a library-grade mechanised repository with common abstractions and coherent APIs \cite{LTEChallenge,CommelinTopazAbstractionBoundaries,BuzzardFLT}.
Building on this success, the Lean Computer Science Library (\cslib{}) is a burgeoning library aimed at providing an infrastructure for the formal development of computer science research, models, and verified software~\cite{cslib_vision}.

In this paper, we introduce the first foundational components and principles in the development and operation of \cslib.
Our contributions include:
\begin{enumerate}
\item The first common abstractions of \cslib, including operational semantics and facilitating computable definitions.
\item A Continuous Integration (CI) pipeline to help with maintenance and handle alignment with Mathlib, which \cslib builds upon.
\item A first-class automation approach to speed up the development of the library.
\item Extensive developments of programming language foundations, aimed both at providing a useful reference and a basis for future extensions.
Notably, these include program equivalences (bisimulation, simulation, and trace equivalence), the Calculus of Communication Systems (a reference example for concurrency theory~\cite{Milner80}), and $\lambda$-calculi with polymorphism and subtyping.
\end{enumerate}

Our work is designed to be a supporting spine for \cslib's ambitious roadmap, which includes a universal formalisation of computer science concepts, verified algorithms, and a verification infrastructure for existing and new programming languages~\cite{cslib_vision}.
Much of this depends on the availability of foundational models and the construction of deeply-embedded languages for reasoning about different aspects of code with mathematical guarantees, which requires APIs for operational semantics and integration with Mathlib.
The infrastructure presented here has already been used in several other developments, including new formalisations of automata theory, Hennessy--Milner Logic, combinatory logic, linear logic, algorithms with time complexity bounds, and more. We see this as an encouraging sign that \cslib{} offers a promising technical platform.

\paragraph{Code and reproducibility.} All code is available at the \cslib public repository\footnote{\href{https://github.com/leanprover/cslib}{github.com/leanprover/cslib}, commit f0d87b6e107dddfe4959dcd2f44ba985e3f1c0d2} and checked with Lean version 4.28.0-rc1.

\section{Shared Abstractions}

\cslib aims to offer reusable abstractions that cut across different areas. We describe some of the first abstractions we developed, which range from  foundational definitions to typeclasses and metaprograms for more overarching connections.

\subsection{Operational Semantics}

Two influential approaches to operational semantics are \emph{abstract reduction systems} \cite{baader_term_1998} and \emph{labelled transition systems} (LTSs) \cite{Sangiorgi_2011}.
The former are captured by binary relations, whereas the latter are relations further parametrised by a type of transition labels.
We similarly bundle both inside structures, as an indication of an API layer enforced by Lean's rules of definitional equality.

\begin{lstlisting}
structure ReductionSystem (Term : Type u) where
  Red : Term → Term → Prop  
\end{lstlisting}

\begin{lstlisting}
structure LTS (State : Type u) (Label : Type v) where
  Tr : State → Label → State → Prop
\end{lstlisting}
We leverage this bundling to automatically derive many standard notions for all instances, like multistep reduction/transitions, reachable states, the image of a state under a transition label, etc.
We also provide a number of standardised propositions and typeclasses to classify systems according to different criteria such as finiteness, image-finiteness, confluence, the diamond property, etc.
These concepts are used consistently across the languages and models of \cslib.

Users can use an attribute to create appropriate arrow notations for single and multistep reductions/transitions, which can be tagged with a user-provided symbol.
These notations work seamlessly with Lean's \lstinline|calc| tactic to facilitate the composition of reduction/transitions in proofs, as demonstrated below.

\begin{lstlisting}[belowskip=0pt]
@[reduction_sys pred_rs "ₙ"]
def Red (a b : ℕ) := a = b + 1 -- simple reductions to predecessors

example : 4 ↠ₙ 1 := by
  calc -- combine reductions to show a multistep reduction from 4 to 1
    4 ⭢ₙ 3 := rfl -- a reduction from 4 to 3
    _ ↠ₙ 1 := by grind -- a multistep reduction from 3 to 1
\end{lstlisting}

\subsection{Computable Definitions}

The Lean ecosystem encourages classical reasoning in several ways, ranging from definitional proof irrelevance to considering choice a `standard axiom' allowed in Mathlib. Definitions that use such reasoning must be marked as \emph{noncomputable}, and Lean does not produce executable code for them. In contrast, we adopt the principle of prioritising computable definitions in \cslib{}. This can include defining computable equivalents to existing definitions in Mathlib.

As an example, \lstinline{Infinite α} characterises the type \lstinline{α} as having no bijection to a finite set. In computer science, it can be useful to  characterise how an infinite type is constructed. With this in mind, we define a class \lstinline{HasFresh α} which is logically equivalent but explicitly carries which carries the data of a function that can be used to construct the type in question. 

\begin{lstlisting}
class HasFresh (α : Type u) : Type (u + 1) where
  fresh : Finset α → α
  fresh_notMem (s : Finset α) : fresh s ∉ s
\end{lstlisting}

We connect these classes with the existing definitions via noncomputable instances, allowing for any proofs (where we likely do not care about computability) to make use of existing theorems from Mathlib.

\subsection{Contexts and Congruences}

Two recurring and associated concepts in language theory are \emph{context} -- terms missing a piece -- and \emph{congruence} -- relations preserved by term constructors.
We offer corresponding typeclasses and a method of registering them as canonical.
For example, the class \lstinline|HasContext| deals with the typical case of a language with a canonical notion of single-hole context. Canonicity is codified by bundling the type of contexts, so Lean can find an instance from knowing the type of terms.
\begin{lstlisting}
class HasContext (Term : Sort*) where
  Context : Sort*
  fill (c : Context) (t : Term) : Term
\end{lstlisting}
Instantiating \lstinline{HasContext} provides the standard notation \lstinline|c[t]| for \lstinline|c.fill t|.
We can then give a general definition of congruence: an equivalence relation covariant (using \lstinline|CovariantClass| from Mathlib) with the \lstinline|fill| operation.
\begin{lstlisting}[belowskip=0pt]
class Congruence (α : Type*) [HasContext α] (r : α → α → Prop) extends
  IsEquiv α r, CovariantClass (HasContext.Context α) α ($\cdot$[$\cdot$]) r
\end{lstlisting}

\section{Proof Automation}

We consider automation a first-class citizen in \cslib:
proof automation is not an afterthought but rather a design constraint on interfaces and definitions. Rather than fixing a formalisation and then attempting to automate it locally, we choose APIs and theorem statements so that routine obligations are discharged by a small set of efficient, generic tactics. In this workflow, failures of automation are treated as feedback on library architecture rather than as isolated proof failures: if a concept repeatedly requires bespoke reasoning, this is a design smell indicating that an interface is poorly factored, a definition is misaligned with common proof patterns, or key normal-form lemmas are missing. This perspective has a compounding effect at library scale: improvements to automation and its supporting `database' of lemma propagate across developments, reducing boilerplate and increasing the stability of reusable infrastructure.

\cslib's automation is primarily based on Lean's \grind, a tactic inspired by SMT solvers that automatically constructs proofs. \cslib has used \grind from nearly its inception, with 314 of 338 declarations using \grind doing so from their initial commit. While this makes measurement of improvement more difficult, the remainder on average have saved 7.1 LoC per theorem upon adding \grind. Of particular note are proofs involving bisimilarity, which had the largest savings ranging from 15 to 39 LoC. As an additional data point, our System F formalization, closely modelled after a Rocq development itself using proof automation techniques, measures at approximately 45\% fewer LoC. While there is subtlety in measuring trade-offs in performance, readability, and proof stability, we are encouraged by these early results.  

As an example of what our implementation can do, even non-trivial results about automata that accept finite strings can be quickly resolved by \grind in \cslib.
Consider the standard subset construction to translate an NFA (nondeterministic finite-state automaton) into a DFA (deterministic finite-state automaton) \cite{Hopcroft2006}. An NFA is just a special case of an LTS in \cslib, so we can reuse all definitions from the latter.
In particular, the next definition returns the set of all states reachable by (any state in) a set of states.
\begin{lstlisting}
def LTS.setImage (S : Set State) (μ : Label) := $\bigcup$ s ∈ S, lts.image s μ
\end{lstlisting}
Given an NFA \lstinline|nfa|, the transition function for the corresponding DFA is simply \lstinline|nfa.setImage|. Thanks to all the theorems that \grind knows about from LTS regarding images, proving that the languages of an NFA and its translation as a DFA are the same has a very simple proof (below given for the general case of automata with potentially-infinite states).
\begin{lstlisting}[belowskip=0pt]
theorem toDAFinAcc_language_eq {na : NA State Symbol} :
    language na.toDAFinAcc = language na := by ext xs; grind
\end{lstlisting}
\section{Reusing Infrastructure}

\cslib has Mathlib, with its over two million lines of formalised mathematics, as a dependency. With nearly a decade of experience to build upon, \cslib has developed a philosophy of reusing existing Mathlib infrastructure whenever reasonable.
This applies not only to reusing definitions and theorems, but also continuous integration, testing, and code quality analysis.

\subsection{Linters}

Lean 4 is self-hosted \cite{moura_lean_2021} and provides a rich system of metaprogramming and hygienic macros \cite{ullrich_extensible_2023}. This supported the growth of a rich ecosystem of linters \cite{baanen_growing_2026,van_doorn_maintaining_2020}, especially in Mathlib, which \cslib reuses and configures. \cslib also extends this set of linters, for example enforcing stricter namespacing policies.

Our usage of linters is similar to Mathlib. \emph{Syntax linters} provide immediate warnings to users which are escalated to errors in CI, while slower \emph{environment linters} that require a completed build are only run in CI. Additional linters with more considerable performance impact are run only weekly, with results reported to the \cslib community channel on the Lean Zulip instance. Similarly to Mathlib, our approach to proof automation prompted us to adopt the expensive \lstinline{mergeWithGrind} linter in this weekly testing: for each use of the \grind tactic, this linter checks if the preceding tactic is redundant.

\subsection{Testing}

In addition to linters and compiler checks, \cslib has a number of standalone tests. These include several examples for \cslib modules, which test also the behaviour of elaborators and import compatibility with Mathlib. Further, our test suite performs a library-wide analysis of the rules specified to \grind, checking for any combination of annotations that instantiates beyond a given threshold (preventing performance issues). Another particular test has a metaprogram that constructs the import graph of \cslib and checks that all modules transitively include a special initialisation module containing all expected linters and tactics.

Lean use a small trusted kernel satisfying the de Brujin criterion \cite{10.1098/rsta.2005.1650}. To mitigate risk even further, one can export kernel declarations for an entire library and have it typechecked by other implementations.
\cslib is periodically analysed in this way through the \emph{Lean Kernel Arena}\footnote{https://arena.lean-lang.org/}, a platform maintained by the Lean FRO that collects the results of these external checkers.

\section{Major Developments Enabled by \cslib}

We now present some of the first major technical developments enabled by \cslib's infrastructure.
The aim of these developments is threefold: to validate and offer feedback to our shared abstractions; to provide general libraries for downstream projects interested in their topics (e.g., concurrency, model checking, and computability theory); and to offer reference examples for contributors.

\subsection{Behavioural Equivalences and CCS}

Building on LTS, we have developed an extensive formalisation of program equivalences, \cslib's first process calculus, and their metatheory.

\subsubsection{Bisimilarity, Similarity, and Trace Equivalence}
We provide bisimilarity, weak bisimilarity, simulation equivalence, similarity, and trace equivalence \cite{Sangiorgi_2011}.
All are parametric in the LTS being considered and are proven to be equivalence relations. Bisimilarity is proven to be strictly finer than trace equivalence.

An important result is that the union of two bisimulations is a bisimulation. We use this and other results to prove that bisimulations and union form a join-semilattice with bottom and top elements -- respectively the empty relation and bisimilarity itself. These results integrate with Lean's and Mathlib's order APIs.

We also prove sound two well-known helper methods \cite{Sangiorgi_2011}:
the bisimulation up-to method, which allows for stopping the bisimulation game as soon as two bisimilar states are reached; and the sw-bisimulation method, which simplifies the weak bisimulation game so that the initial challenge is a single transition.

\subsubsection{Calculus of Communicating Systems (CCS)}
We formalise Milner's CCS \cite{Milner80}, as given in the textbook presentation of bisimulation and its algebraic methods \cite{Sangiorgi_2011} (expressive enough to study concurrency patterns like semaphores).

The language of CCS processes is parametric on both the types of names and constants (procedures).
We instantiate \lstinline|HasContext| for CCS and prove that it completely captures all process contexts.
\begin{lstlisting}
theorem Context.complete (p : Process Name Constant) : ∃ c : Context _ _,
    p = c[Process.nil] ∨ ∃ k : Constant, p = c[Process.const k]
\end{lstlisting}

We then use our bisimilarity module to prove well-known behavioural laws of CCS such as commutativity and distributivity of parallel composition. Our development culminates in the hallmark result that bisimilarity is a congruence for CCS, regardless of the concrete definitions of constants (\lstinline|d| below).
\begin{lstlisting}[belowskip=0em]
instance bisimilarityCongruence :
    Congruence (Process Name Constant) (Bisimilarity (lts (defs := d)))
\end{lstlisting}

\subsection{Lambda Calculi and Binding Metatheory}

\cslib contains a growing formalisation of the metatheory of $\lambda$-calculi, currently including the simply typed $\lambda$-calculus and System F with subtyping. While this is standard foundational work, our use of locally nameless variable binding \cite{chargueraud_locally_2012} in conjunction with proof automation and metaprogramming is novel.

\subsubsection{Typing Contexts}

Locally nameless typing contexts are indicative of the \cslib philosophies of reusing and extending Mathlib infrastructure. In this representation of binding, a typing context is simply a list of pairs of free variables and their type ascription. Mathlib has a module \lstinline{Mathlib.Data.List.Sigma} that matches this purpose and we are able to almost completely reuse. The only additional definitions are for notational convenience, and \cslib inherits \grind annotations from Mathlib. Notably, we add an additional set of \grind rules that is scoped to \cslib's namespace of locally nameless typing contexts. In this way we encapsulate aggressive automation that may not be more widely appropriate.

\subsubsection{Free Variable Section as Term Elaboration}

A common feature of languages with binders is selection of fresh variables with respect to terms, types, and contexts. While prior work has provided tactics for locally nameless variable selection, these require specialization for each type system and lack user configurability. Instead of a tactic we define a \emph{term elaborator}, a metaprogram that expands user syntax into a term. Elaboration into ordinary terms aids compatibility with typeclass inference (in particular \lstinline{HasFresh}) and proof automation.

As shown below, the \lstinline{free_union} elaborator selects all free variables in the local context using any number of user-specified functions. During elaboration these functions are applied to any appropriate hypothesis and assembled into a finite set of variables (including by default variables and sets of variables). 

\begin{lstlisting}[belowskip=0pt]
variable (x : ℕ) (xs : Finset ℕ) (var : String)
def f (_ : String) : Finset ℕ := {1, 2, 3}

#check free_union [f] ℕ -- info: ∅ ∪ {x} ∪ id xs ∪ f var : Finset ℕ
\end{lstlisting}

\section{Related Work and Conclusion}

While there have been significant formalisation efforts in Lean within the realm of computer science such as SampCert \cite{SampCert} and Cedar \cite{cutler_cedar_2024}, there has up to this point not been a monolithic library of computer science in Lean as envisioned by \cslib \cite{cslib_vision}, offering a unified platform with cross-cutting abstractions and development principles.
In this article, we have presented the first technical steps into making this vision a reality.

There is much room to extend \cslib's treatments of binding and reasoning on LTSs; we give a few examples. Building on prior work in Rocq \cite{schafer_autosubst_2015,stark_autosubst_2019,dapprich_generating_2021,aydemir_lngen_2010} and Isabelle \cite{urban_nominal_2008,urban_formal_2005}, we are developing an interface that will allow for specifying type systems fully within Lean and include a configurable metaprogramming engine for generating definitions for binding and substitution.
Inspired by formalisations of nominal transition systems \cite{PBEGW21}, we are developing logics for reasoning about LTSs where transitions can bind names into derivatives -- as in the $\pi$-calculus \cite{MPW92a}.
Looking beyond the scope of what has been formalised in the literature so far, we aim at leveraging our APIs for capturing and reasoning about algorithms, database theory, distributed systems, and more.

\subsubsection*{Acknowledgements}

We thank the online communities of \cslib, Lean, and Mathlib for useful discussions. In particular the Lean FRO has been invaluable in their early technical assistance.

Co-funded by the European Union (ERC, CHORDS, 101124225). Views and opinions expressed are however those of the authors only and do not necessarily reflect those of the European Union or the European Research Council. Neither the European Union nor the granting authority can be held responsible for them. This work was supported in part by the US National Science Foundation under Grant No.: \href{https://www.nsf.gov/awardsearch/showAward?AWD_ID=2220991}{CCF-2220991}.

\nocite{*}
\bibliographystyle{eptcs}
\bibliography{references}
\end{document}